\documentclass[10pt]{iopart}

\usepackage{iopams}
\usepackage[utf8]{inputenc}
\usepackage{graphicx}
\usepackage{color}
\usepackage{overpic}
\usepackage{siunitx}
\usepackage{soul}

\usepackage{cite}
\usepackage{hyperref}
	\hypersetup{colorlinks=true,
		pdfstartview=FitV,
		linkcolor=blue,
		citecolor=red,
		urlcolor=blue,
		pdfauthor={HOPPE Mathias, hoppe@chalmers.se},
		pdfsubject={Synchrotron in a vertically translated plasma},
		pdftitle={Synchrotron in a vertically translated plasma}
	}

\newcommand{\eqref}[1]{(\ref{#1})}

\newcommand{\figlabelw}[1]{\textcolor{white}{\textbf{#1}}}

\newcommand{\fRE}{f_{\rm RE}}
\newcommand{\LIUQE}{\textsc{Liuqe}}
\newcommand{\MultiCam}{MultiCam}
\newcommand{\SOFT}{\textsc{Soft}}
\newcommand{\Pwr}{\mathcal{P}}

\newcommand{\pstar}{p^\star}
\newcommand{\tstar}{\theta^\star}

\begin{document}
    \title[Runaway synchrotron in vertically translated plasma]{Runaway electron synchrotron radiation in a vertically translated plasma}
    \author{M~Hoppe$^1$, G~Papp$^2$, T~Wijkamp$^{3,4}$, A~Perek$^3$, J~Decker$^5$, B~Duval$^5$, O~Embreus$^1$, T~F\"ul\"op$^1$, U~A~Sheikh$^5$ the TCV Team$^\dagger$ and the EUROfusion MST1 Team$^*$}
    \address{
        $^1$Department of Physics, Chalmers University of Technology, SE-41296 Gothenburg, Sweden\\
        $^2$Max Planck Institute for Plasma Physics, D-85748 Garching, Germany\\
        $^3$FOM Institute DIFFER `Dutch Institute for Fundamental Energy Research', 5600 HH Eindhoven, Netherlands\\
        $^4$Department of Applied Physics, Eindhoven University of Technology, Eindhoven 5600 MB, Netherlands\\
        $^5$Swiss Plasma Centre, EPFL, CH-1015 Lausanne, Switzerland\\
        $^\dagger$See the author list of ``S. Coda  {\em et al.}, 2019 Nucl. Fusion \href{https://iopscience.iop.org/article/10.1088/1741-4326/ab25cb}{{\bf 59} 112023}''\\
        $^*$See the author list of ``B. Labit {\em et al.}, 2019 Nucl. Fusion \href{https://iopscience.iop.org/article/10.1088/1741-4326/ab2211}{{\bf 59} 086020}''\\
    }
    \ead{hoppe@chalmers.se}
    \vspace{10pt}
    \begin{indented}
        \item[]February 2020
    \end{indented}
    
    \begin{abstract}
      Synchrotron radiation observed from runaway electrons (REs) in
      tokamaks depends upon the position and size of the
      RE beam, the RE energy and pitch distributions, as well
      as the location of the observer.
      We show experimental synchrotron images of a vertically moving runaway
      electron beam sweeping past the detector in the TCV tokamak and compare it
      with predictions from the synthetic synchrotron diagnostic \SOFT. This
      experimental validation lends confidence to the theory
      underlying the synthetic diagnostics which are used for
      benchmarking theoretical models of and probing runaway dynamics.
      We present a comparison of synchrotron measurements in TCV with
      predictions of kinetic theory for runaway dynamics in uniform
      magnetic fields. We find that to explain the detected
      synchrotron emission, significant non-collisional pitch angle
      scattering as well as radial transport of REs would be needed.
      Such effects could be caused by the presence of magnetic
      perturbations, which should be further investigated in future
      TCV experiments.
    \end{abstract}
    
    \submitto{\NF}
    
    \ioptwocol
    
    \section{Introduction}\label{sec:intro}
    One of the key issues facing future reactor-scale tokamaks, such
    as ITER, is plasma terminating disruptions. Such events can
    convert a significant part of the plasma current to
    relativistic runaway electron (RE) current. Uncontrolled loss of
    the RE current may then damage the plasma facing components and must
    be avoided. Therefore, in recent years, much effort has been
    devoted to developing strategies for preventing or mitigating the
    effect of REs \cite{Lehnen2015,Hollmann2015,Breizman2019review}. Evaluating these
    strategies requires reliable theoretical models for RE generation
    and their subsequent dynamics. However, most theoretical and
    numerical models currently available to the community, and used
    e.g.\ in predictions for ITER, have yet to be thoroughly benchmarked
    against experimental data.

    A powerful diagnostic method for REs within
    a plasma is to measure their {\em synchrotron radiation} \cite{Jaspers2001}.
    The radiation spectrum emitted by the runaways, and the associated
    radiation spot observed in camera images, both depend on the
    momentum- and real-space distribution of the runaways.
    Synchrotron radiation measurements therefore provide
    insight into the pitch angle,
    energy~\cite{PazSoldan2018,Tinguely2018a}
    and spatial distribution~\cite{Tinguely2018b} of runaway electrons.

    Measurements of synchrotron radiation emitted by REs have been
    performed on tokamaks since the early
    90's~\cite{Finken1990,Jaspers1995}, with both
    visible-light/infrared spectrometers and cameras.  During recent
    years, more advanced synthetic diagnostic tools have been
    developed~\cite{Carbajal2017,Hoppe2018a} that take into account
    the camera position and magnetic equilibrium.
    The application of synthetic synchrotron diagnostics to bridge the
    gap between theory and experiment is still under development,
    and the applicability of the synthetic diagnostics are not yet
    fully experimentally validated.

    This letter reports a recent runaway electron
    experiment conducted at the Tokamak \`a Configuration Variable
    (TCV), situated at the Swiss Plasma Center in Lausanne,
    Switzerland.
    The highly elongated shape of the TCV chamber, combined with its rich
    set of poloidal field coils and sophisticated control
    software~\cite{Carnevale2019,Decker2020}, provides a unique opportunity
    to investigate the vertical dependence of synchrotron radiation emission.
    In TCV a high current conversion, fully developed runaway electron beam can
    be reliably displaced vertically over a distance comparable to the minor
    radius~\cite{Carnevale2019}.
    In the present experiment, a runaway electron beam was generated and maintained
    within an ohmically driven plasma in a non-disruptive phase. The circular,
    limited plasma is displaced vertically within the vacuum chamber, sweeping
    across the synchrotron camera's field-of-view and probing how the
    synchrotron spot depends upon the relative vertical position of the
    runaway beam and synchrotron detector. The experimental measurements show
    good qualitative agreement with the predictions obtained from the synthetic
    diagnostic tool \SOFT~\cite{Hoppe2018a}. This experiment complements
    previous publications~\cite{Tinguely2018a,Tinguely2018b,Tinguely2019}
    by providing proof of the \emph{vertical position} spot shape predictions
    from the synthetic synchrotron diagnostic.

    \section{Synchrotron radiation spot shapes}\label{sec:theory}
    Synchrotron radiation is emitted primarily along the velocity
    vector of the emitting particle\cite{Schwinger1949}.
    Synchrotron radiation, thus, is only observed when the emitting particle
    is moving directly towards the observer. In tokamak plasmas,
    synchrotron emission appears as asymmetric patterns of
    emissivity, the shape of which is sensitive to both the RE
    distribution function, as well as the detector position and
    magnetic field. The dependence of the synchrotron spot shape on
    these parameters has been extensively
    studied~\cite{Carbajal2017,JaspersThesis,Pankratov1996,Zhou2014,Hoppe2018a}.
    In this section, we restrict our attention to the emission shape
    dependence on the vertical distance $\Delta Z$ between the detector and
    runaway beam.

    Often, the observed synchrotron spot depends sensitively only on a
    small subset of the runaway electron distribution function $\fRE$,
    namely where the product
    $\Pwr(r,p,\theta) = G(r,p,\theta)\fRE(r,p,\theta)$ attains a
    maximum. Here, $G(r,p,\theta)$ denotes the radiated
    power recorded by a given detector from an ensemble of electrons
    located on the same flux-surface labelled by $r$ with the same momentum
    $p$ and pitch angle $\theta$ in the point of
    minimum magnetic field along the orbit. If $\Pwr$ is sharply
    peaked at the point $(\pstar, \tstar)$, the synchrotron spot will
    have nearly the same shape as if all particles had the same
    momentum and pitch angle
\begin{equation}\label{eq:fREdelta}
    \fRE\sim\delta(p-\pstar)\delta(\cos\theta-\cos\tstar).
\end{equation}
It is therefore often possible to characterise synchrotron spots by
the pair $(\pstar,\tstar)$ of parameters corresponding to the maximum
of $\mathcal{P}$, i.e.\ the subset of runaway electrons that dominate
synchrotron emission~\cite{Hoppe2018b}. While this is most likely far from the
true distribution of runaway electrons, this approximation matches
experimental synchrotron measurements sufficiently well for the TCV discharge
analysed in this letter. Other, physically motivated models for the
distribution function were also considered for the TCV discharge, but for
reasons to be discussed in section~\ref{sec:experiment}, these failed to
predict the observed synchrotron radiation pattern.
Therefore, in what follows, we will only
consider synchrotron spots resulting from distribution functions of
the form in equation~\eqref{eq:fREdelta}.

To simulate the detected synchrotron radiation, we use the open source
synthetic synchrotron diagnostic \SOFT~\cite{Hoppe2018a}. \SOFT\ 
calculates the intensity of bremsstrahlung and/or synchrotron radiation
reaching a given detector, in an arbitrary, axisymmetric magnetic
geometry. The guiding-center formulation used in \SOFT\ ensures short
simulation times, while accurately accounting for the magnetic geometry,
orbit drifts and radiation spectrum dependence on geometrical and particle
parameters.

% Vertical dependence
The position of the synchrotron detector relative to the plasma
strongly affects the observed radiation spectrum and spot shape and,
in particular, the vertical detector placement can change the observed
part of the distribution function, i.e.\ the values of $\pstar$ and $\tstar$.
If the vertical offset $\Delta Z$ of the detector relative to the plasma is
non-zero, then electrons located within a radius
\begin{equation}\label{eq:rlim}
    r \lesssim \frac{\Delta Z - D\tan\theta}{1+(D+\Delta Z\tan\theta)\iota/2\pi R_{\rm m}},
\end{equation}
are not observed,
where $\iota$ is the rotational transform, $R_{\rm m}$ the tokamak major
radius and $D$ the distance between the detector and runaway beam, since
the electron velocity vectors never intersect the
detector aperture. This may alter the observed part of the distribution
function, possibly favouring the observation of runaways with
large pitch angles. In practice, however, a vertical detector offset
most likely always has a limited effect on the observed part of the distribution,
since the total synchrotron radiation emitted by a single particle
scales as $\sim\sin^2\theta$, inherently favouring the
detection of large pitch-angle particles. For TCV discharge \#64614,
equation~\eqref{eq:rlim} yields a threshold pitch angle
$\tan\theta_{\rm th}\lesssim\Delta Z/D\sim 0.1$, meaning that particles
with smaller pitch angles could be out of the camera view. Hence,
if the pitch angle $\tstar$ of the dominant particles is close to or
smaller than $\theta_{\rm th}$, one could see a significant dependence
on $\Delta Z$ in the dominant parameters $\pstar$ and $\tstar$.
Conversely, if $\tstar>\theta_{\rm th}$, the dependence
on $\Delta Z$ is weak for both $\pstar$ and $\tstar$.

% Benefits:
A potential benefit of multiple camera views at different
vertical positions was pointed out in
ref.~\cite{JaspersThesis}. There, it was shown, albeit in a simplified
geometry, that it is not possible to differentiate between a
synchrotron spot resulting from runaways with an intermediate
($\sin\theta\sim r_{\rm beam}/D$) against a
large pitch angle ($\sin\theta\gtrsim r_{\rm beam}/D$) using a single
synchrotron camera. Using two cameras situated at
different vertical positions, this degeneracy can be broken, and
a wider range of pitch angles may be differentiated.

This conclusion holds for a more advanced treatment,
such as the \SOFT\ simulations, although the reasons differ.
With \SOFT, it is still possible to differentiate
between pitch angles due to the pitch angle dependence of
the toroidal origin of the radiation. Synchrotron spots corresponding
to smaller pitch angles, however, tend to have distinct bright areas,
running along the upper and lower edges of the spot~\cite{Hoppe2018b}.
These disappear for sufficiently high pitch angles, but by
using two cameras at different vertical positions, it should be
possible to extend the range of pitch angles for which the bright
areas can be observed. This would, in particular, be useful when
studying the (horizontal) polarization of synchrotron radiation that
further emphasises the bright areas of the synchrotron
spot~\cite{Hoppe2018eps,Tinguely2019}.
The additional information gained from having two
vertically offset cameras is in most cases limited, compared to
other possible system upgrades (e.g.\ observing the radiation in
multiple wavelength ranges, combining camera data with spectral
measurements, or measuring the polarisation).

\section{Observation of a vertically translated plasma}\label{sec:experiment}
\begin{figure}
    \centering
    \includegraphics[width=0.49\textwidth]{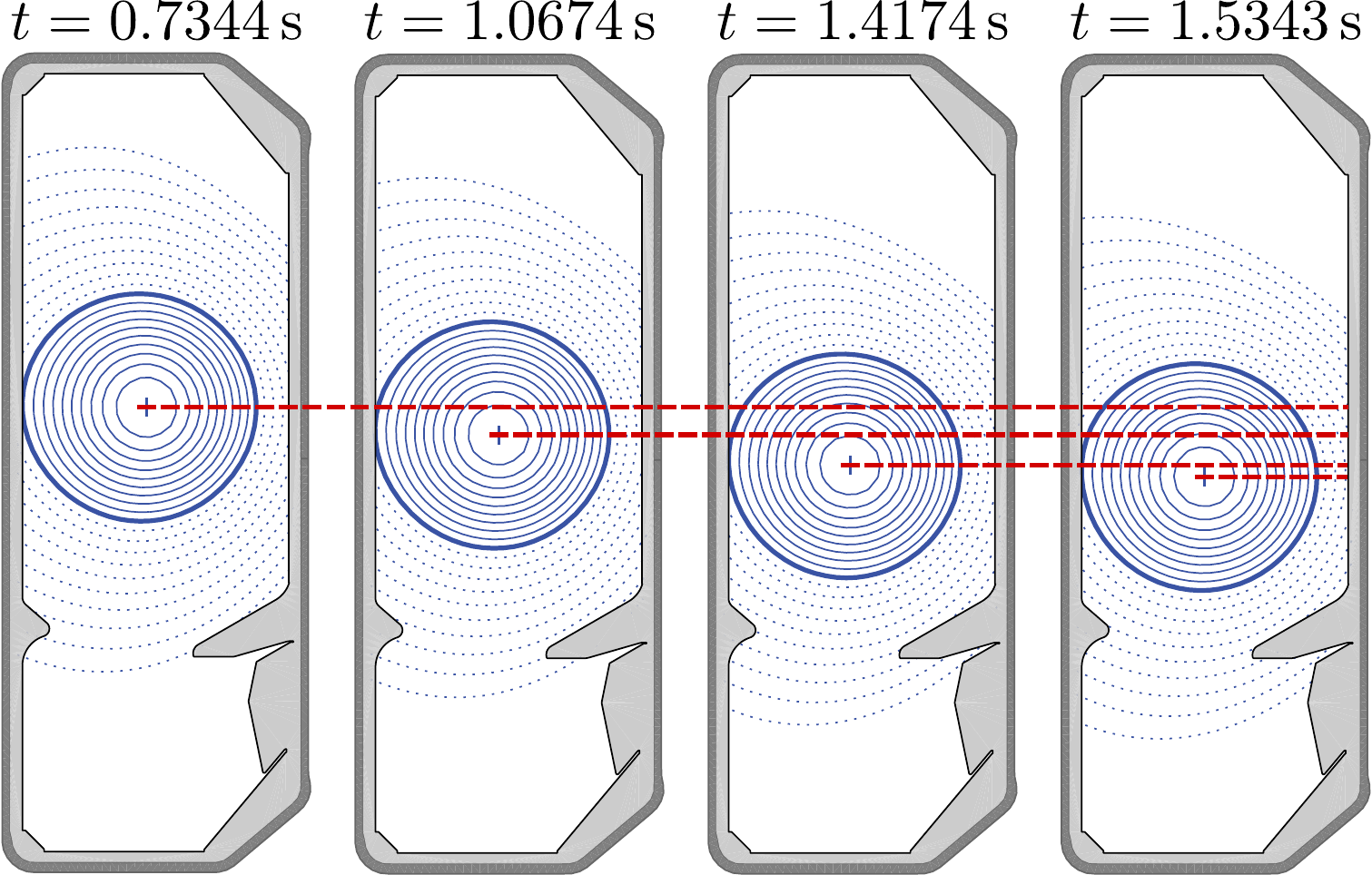}
    \caption{
        TCV plasma cross-section at approximately the four times considered in
        this letter. The plasma downward motion is indicated with the green
        dashed lines. The vertical magnetic axis location $Z$ in these plasmas is,
        in order from earlier to later: $Z=\SI{0.107}{m}$, $Z=\SI{0.057}{m}$,
        $Z=-\SI{0.003}{m}$, and $Z=-\SI{0.020}{m}$.
    }
    \label{fig:TCVplasma}
\end{figure}

\begin{figure*}
    \centering
    \begin{overpic}[width=\textwidth]{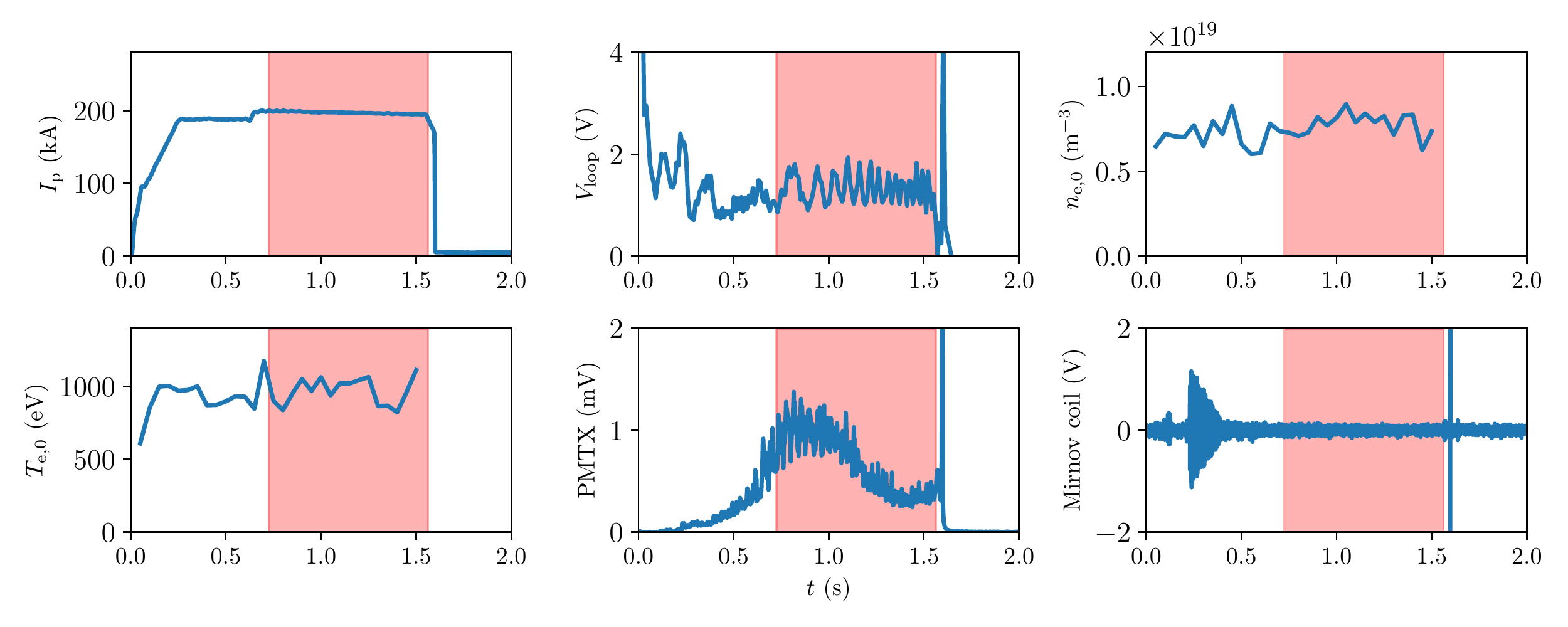}
        \put(9,34.3){(a) Plasma current}
        \put(41,34.3){(b) Loop voltage}
        \put(74,34.3){(c) Electron density}
        \put(9,17){(d) Electron temperature}
        \put(41,17){(e) Hard x-rays}
        \put(74,17){(f) Mirnov coil}
    \end{overpic}
    \caption{
        Time evolution of plasma parameters in TCV \#64614. (a) plasma current,
        (b) loop voltage at plasma edge, (c) core electron density (from Thomson
        scattering), (d) core electron temperature (from Thomson scattering),
        (e) hard x-ray signal, (f) Mirnov coil signal. The timespan during which
        the plasma was vertically translated is indicated by the red shaded
        region.
    }
    \label{fig:64614}
\end{figure*}
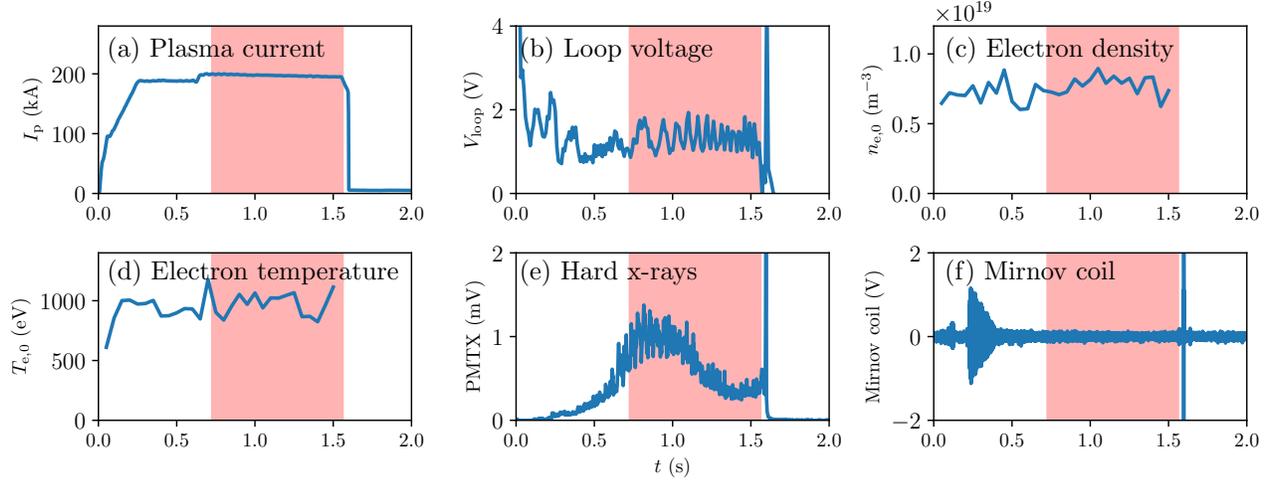

The elongated vacuum chamber cross-section of TCV, illustrated in
figure~\ref{fig:TCVplasma}, combined with excellent plasma control
capabilities~\cite{Carnevale2019,Decker2020}, makes TCV an ideal tokamak for studying the dependence
of the runaway electron synchrotron spot on the vertical distance
between the camera and runaway beam, which has
not been studied experimentally before.
In the approximately 1 second long, quiescent flat-top
of TCV discharge \#64614 (a type of plasma which references~\cite{PazSoldan2014,Esposito2017}
describe the general characteristics of), summarised in figure~\ref{fig:64614},
a population of runaway
electrons was generated and subsequently translated vertically from
$Z = \SI{10.7}{cm}$ to $Z = -\SI{2.0}{cm}$ ($\sim 60\%$ of the plasma minor
radius). Before displacing the plasma, the synchrotron spot was
allowed to develop and reach an asymptotic shape. The
ohmically-heated discharge,
$I_{\rm p} = \SI{200}{kA}$, circular, limited plasma had a toroidal
magnetic field $B_{\rm T} = \SI{1.43}{T}$, major radius
$R_{\rm p}=\SI{0.86}{m}$ and minor radius $a_{\rm p} = \SI{0.21}{m}$.
The core electron density was held constant at
approximately $n_{\rm e} = \SI{0.8d19}{\per\cubic\metre}$, and the core
electron temperature at about $T_{\rm e} = \SI{1}{keV}$. A constant edge
loop voltage was applied, resulting in an estimated electric field of
$E\approx\SI{0.25}{V/m}$ at the plasma center ($E/E_{\rm c}\approx 30$,
$E/E_{\rm D}\approx 6\%$, where $E_{\rm c}$ is the Connor-Hastie field~\cite{Connor1975}
and $E_{\rm D}$ the Dreicer field~\cite{Dreicer1959}), enabling runaway generation.
Visible images were recorded using the multispectral imaging system
\MultiCam\ that distributes incoming light over four channels
with different narrowband filters using beamsplitters so
channels have a nearly identical observation geometry. Camera and
data processing specifications are the same as those for the
\textsc{Mantis} system~\cite{Perek2019}. Figure~\ref{fig:expsynch} shows a
selection of undistorted \MultiCam\ images of the evolution of the
synchrotron spot through a narrowband filter centered at $\SI{640.6}{nm}$
with full-width at half maximum of $\SI{1.73}{nm}$ (this range is selected
for the lack of strong line emission). The detector parameters used
for the simulations are listed in table~\ref{tab:mantisparams}.

\newlength{\tcveqspace}
\setlength{\tcveqspace}{3mm}

\begin{table}
    \centering
    \caption{Parameters for the \MultiCam\ detector used as input to \SOFT\ simulations.}
    \label{tab:mantisparams}
    \begin{tabular}{c|c}
        \textbf{Parameter} & \textbf{Value}\\\hline
        Position $(x,y,z)$ & $(1.082, -0.346, 0.014)\,\si{\metre}$\\
        Viewing direction & $(-0.794, -0.482, 0.054)$\\
        Vision angle & $\SI{0.608}{rad}$\\
        Camera roll & $\SI{0.023}{rad}$ (CCW)\\
        Wavelength & $\SI{640.6}{nm}$
    \end{tabular}
\end{table}

\newlength{\expimgwidth}
\setlength{\expimgwidth}{0.117\textwidth}
\begin{figure}[t!]
    \centering
    \begin{overpic}[width=\expimgwidth]{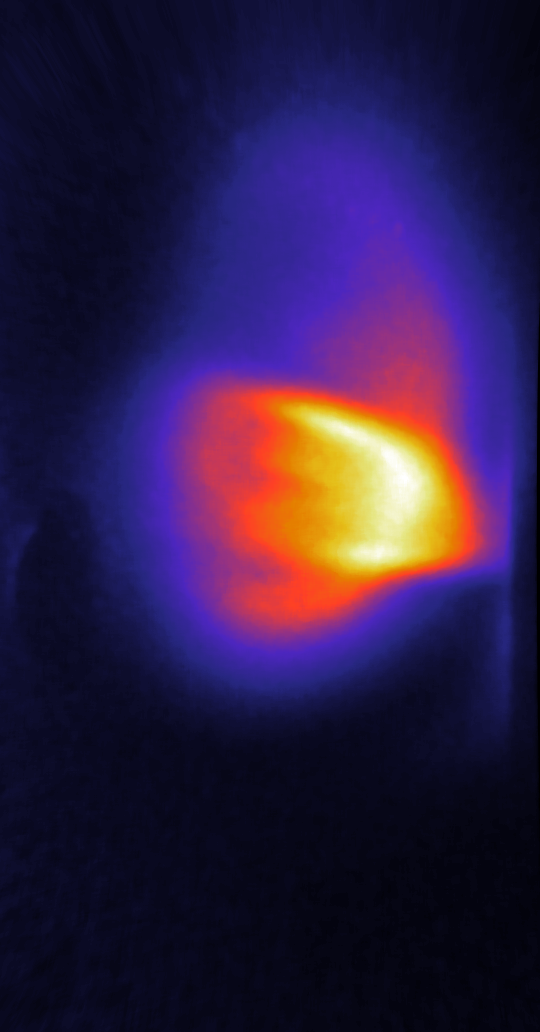}
        \put(4,91){\small\figlabelw{(a)}}
        \put(4,12){\scriptsize\textcolor{white}{$\Delta Z = \SI{9.3}{cm}$}}
        \put(4,5){\scriptsize\figlabelw{$t=\SI{0.725}{s}$}}
    \end{overpic}
    \begin{overpic}[width=\expimgwidth]{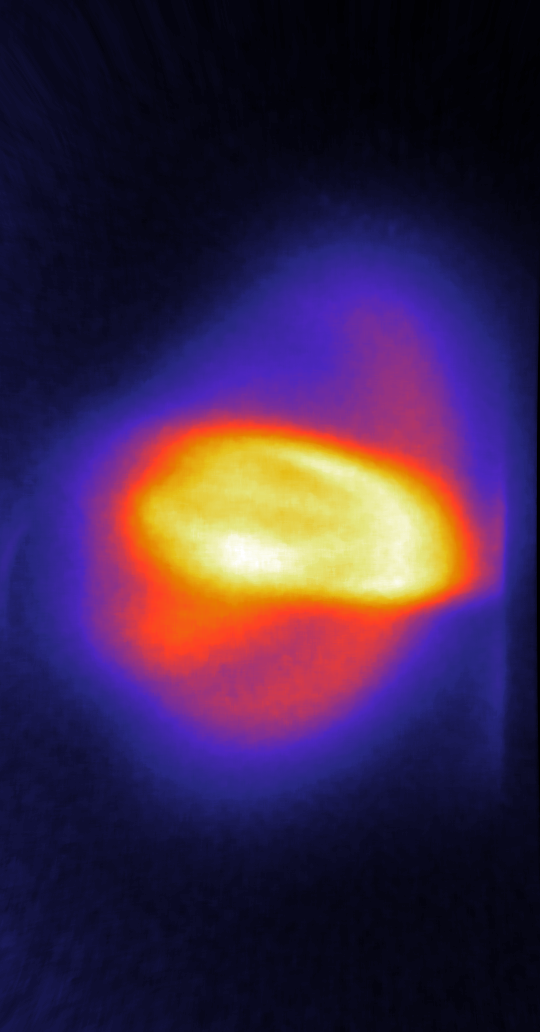}
        \put(4,91){\small\figlabelw{(b)}}
        \put(4,12){\scriptsize\textcolor{white}{$\Delta Z = \SI{4.3}{cm}$}}
        \put(4,5){\scriptsize\figlabelw{$t=\SI{1.069}{s}$}}
    \end{overpic}
    \begin{overpic}[width=\expimgwidth]{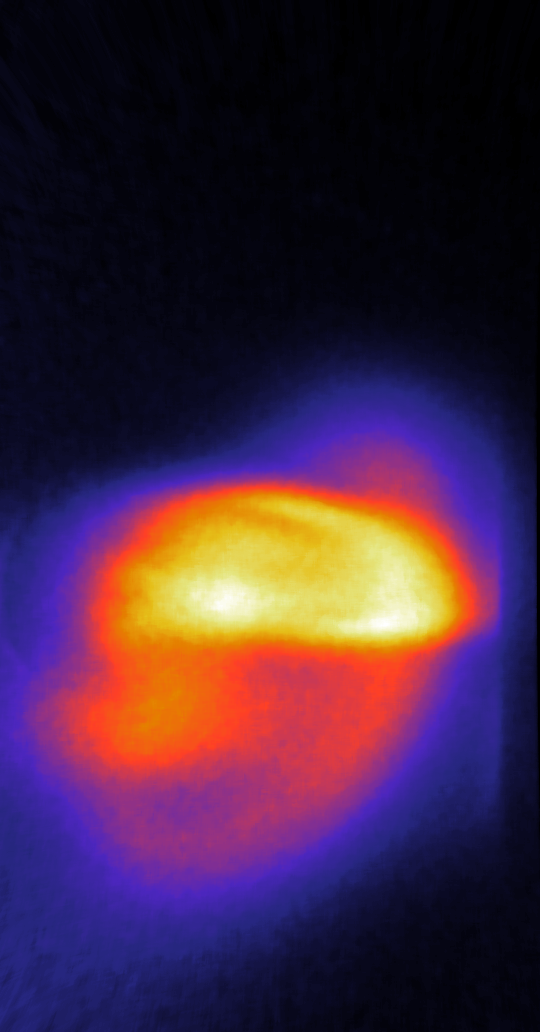}
        \put(4,91){\small\figlabelw{(c)}}
        \put(4,12){\scriptsize\textcolor{white}{$\Delta Z = -\SI{1.7}{cm}$}}
        \put(4,5){\scriptsize\figlabelw{$t=\SI{1.412}{s}$}}
    \end{overpic}
    \begin{overpic}[width=\expimgwidth]{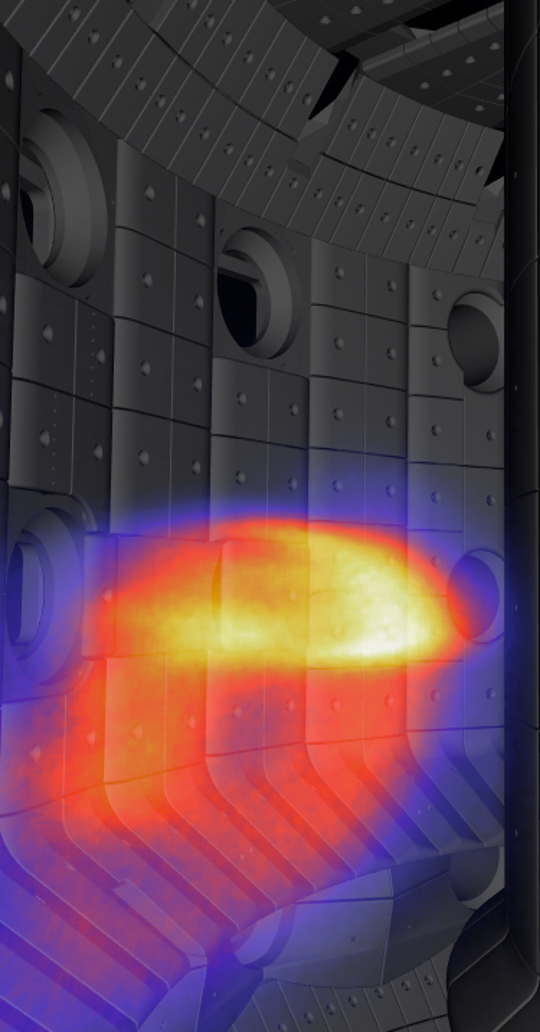}
        \put(4,91){\small\figlabelw{(d)}}
        \put(4,12){\scriptsize\textcolor{white}{$\Delta Z = -\SI{3.4}{cm}$}}
        \put(4,5){\scriptsize\figlabelw{$t=\SI{1.562}{s}$}}
    \end{overpic}
    \includegraphics[width=0.5\textwidth]{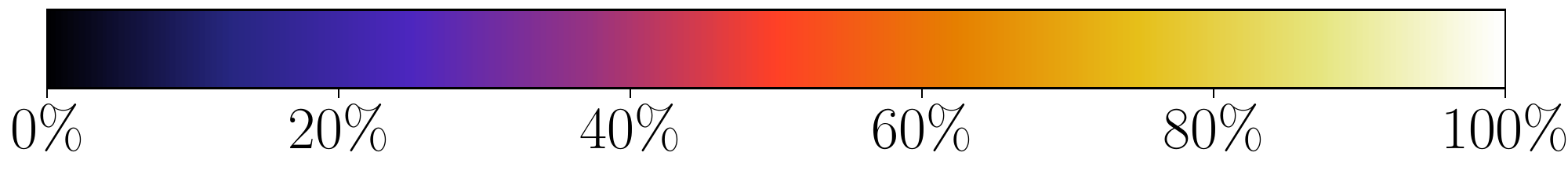}
    \caption{
        Synchrotron radiation images from the \MultiCam\
        camera system during TCV discharge \#64614. The plasma is translated
        vertically downwards, past the camera, and the synchrotron spot
        changes accordingly. A CAD drawing of the camera view has been overlaid
        the image in (d). The two surfaces which make up the synchrotron
        spot can be distinguished and appear to be moving downwards at
        different rates. Note that in these images, the high-field side is
        to the right. Each pixel value indicates the received photon flux
        (photons/s), but due to that the camera was not absolutely calibrated,
        and in order to emphasize the radiation pattern, we normalize each
        frame to the value of its brightest pixel.
    }
    \label{fig:expsynch}
\end{figure}
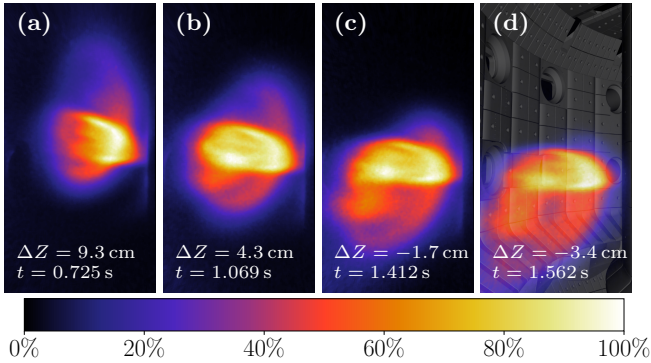

The synchrotron spot in discharge \#64614 consists of two
separate, oval parts---one large, vertically elongated and 
a smaller, horizontally elongated spot. As the plasma is translated
vertically downwards, these two components
correspondingly move downwards in the image, but at different rates.
The smaller spot component (corresponding to runaways far from the
detector) moves only slightly downwards, whereas the larger spot
translates significantly during the scan.

The appearance of two distinct spots suggests that the dominant pitch angle is
relatively large~\cite{Hoppe2018a}. Using \SOFT\ simulations, we compare the
contours of the simulated and experimental synchrotron images,
and estimate the dominant particle parameters to be
$\tstar\approx\SI{0.40}{rad}$ and $\pstar\approx 50m_ec$, where $m_e$ is the electron
rest mass and $c$ the speed of light, so that $m_ec\approx\SI{0.511}{MeV/c}$.
We let the radial density profile of the REs take the form $n_{\rm RE}(r)\propto J_0(x_1r/a_p)$,
where $J_0(x)$ is a zeroth-order Bessel function of the first kind, and $x_1$ its
first zero, which is the steady-state solution assuming a strong, diffusive radial
transport with a uniform diffusion coefficient.
The synchrotron patterns are most sensitive to variations in $\tstar$, and to
support our estimate we present synthetic synchrotron images in the $t=\SI{0.7344}{s}$
magnetic equilibrium for a few
different values of $\tstar$ in figure~\ref{fig:paramscan}. Comparing the
images in figure~\ref{fig:paramscan} to figure~\ref{fig:expsynch}a, we conclude
that $\tstar = \SI{0.4}{rad}$ is close to the lower limit for $\tstar$.
At $\tstar = \SI{0.6}{rad}$, the small synchrotron spot has begun to
disappear behind the tokamak central column, allowing us to conclude that
$\tstar = \SI{0.6}{rad}$ is close to an upper limit for $\tstar$, as no part
of the synchrotron pattern in figure~\ref{fig:expsynch}a is obscured by the
the tokamak wall.
The synchrotron pattern is somewhat less sensitive to $\pstar$,
but the RE energy influences the relative intensities of high-field side and
low-field side emission~\cite{Hoppe2018b}, as well as the shift of the pattern
due to orbit drifts. Based on these observations, we conclude that the dominant
RE momentum should be near $p = 50m_ec$.

\newlength{\scanimgwidth}
\setlength{\scanimgwidth}{0.117\textwidth}
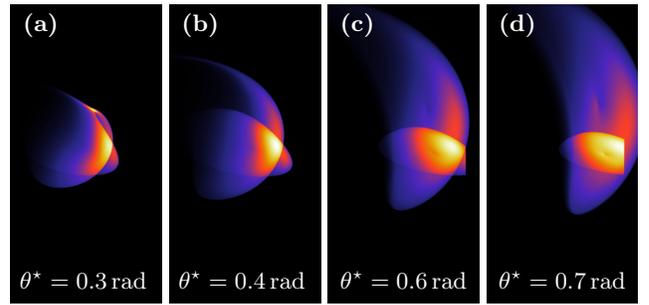
\begin{figure}
    \centering
    \begin{overpic}[width=\scanimgwidth]{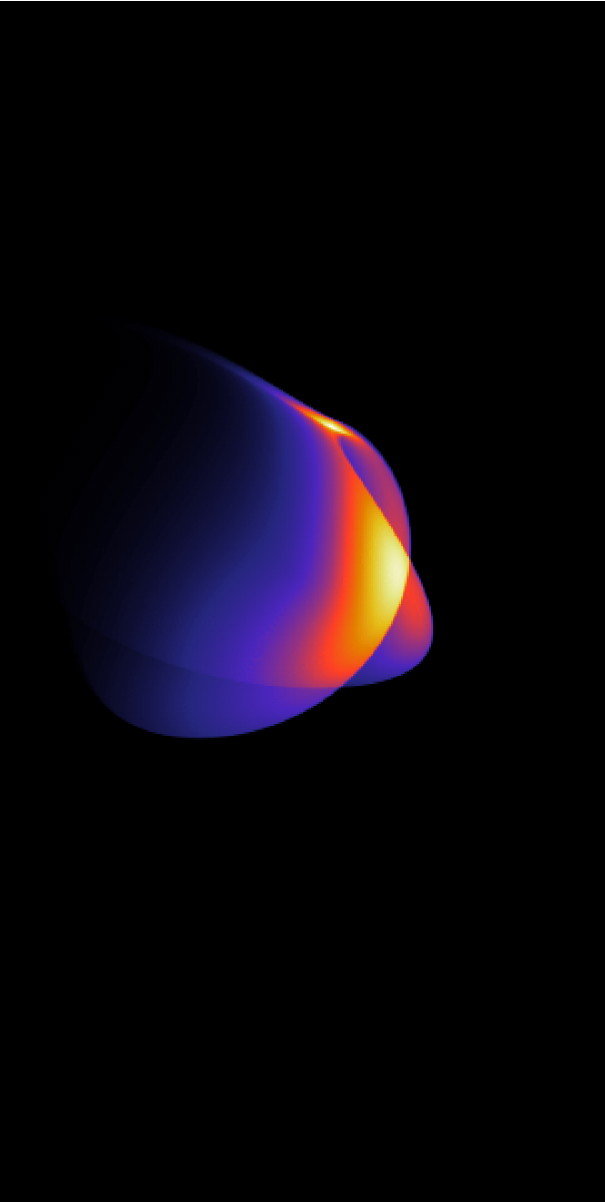}
        \put(4,91){\small\figlabelw{(a)}}
        \put(3,5){\small\figlabelw{$\tstar=\SI{0.3}{rad}$}}
    \end{overpic}
    \begin{overpic}[width=\scanimgwidth]{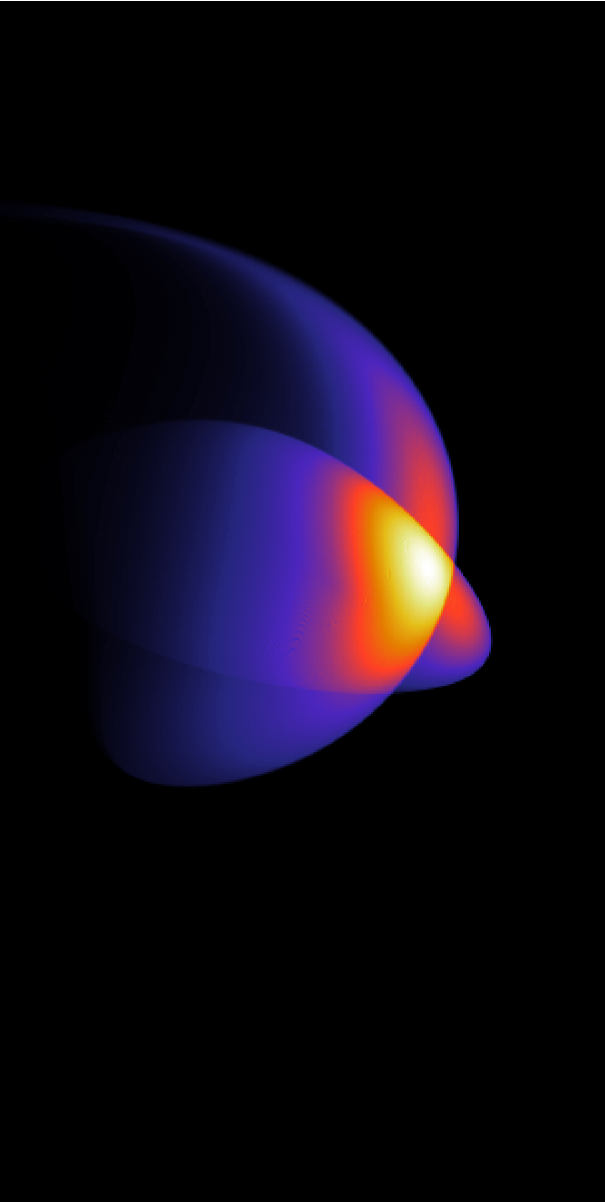}
        \put(4,91){\small\figlabelw{(b)}}
        \put(3,5){\small\figlabelw{$\tstar=\SI{0.4}{rad}$}}
    \end{overpic}
    \begin{overpic}[width=\scanimgwidth]{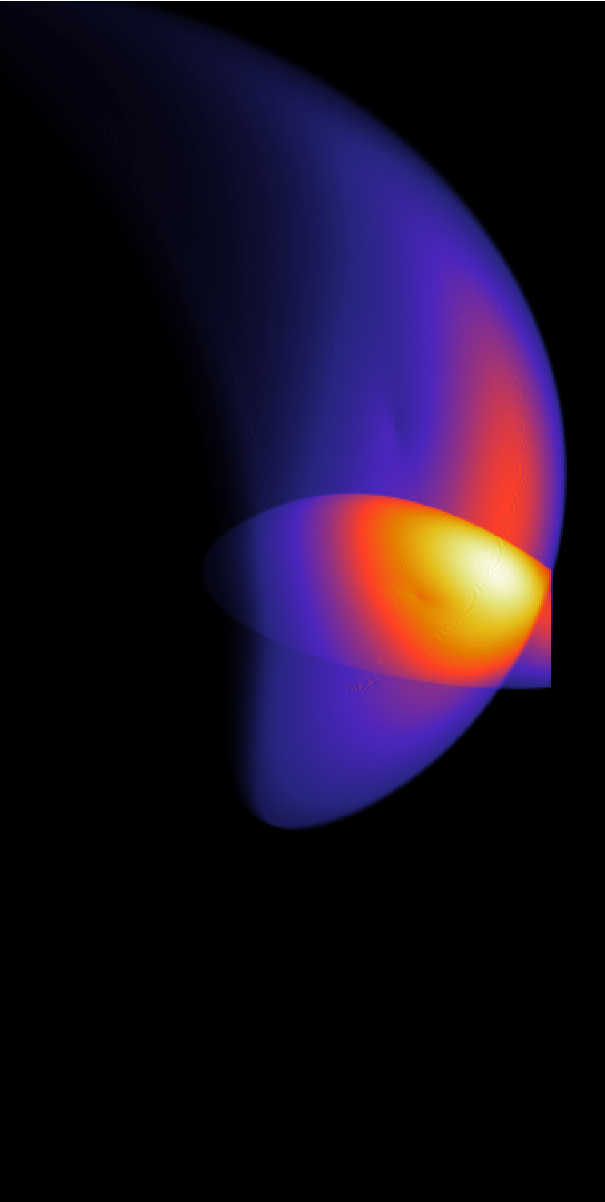}
        \put(4,91){\small\figlabelw{(c)}}
        \put(4,5){\small\figlabelw{$\tstar=\SI{0.6}{rad}$}}
    \end{overpic}
    \begin{overpic}[width=\scanimgwidth]{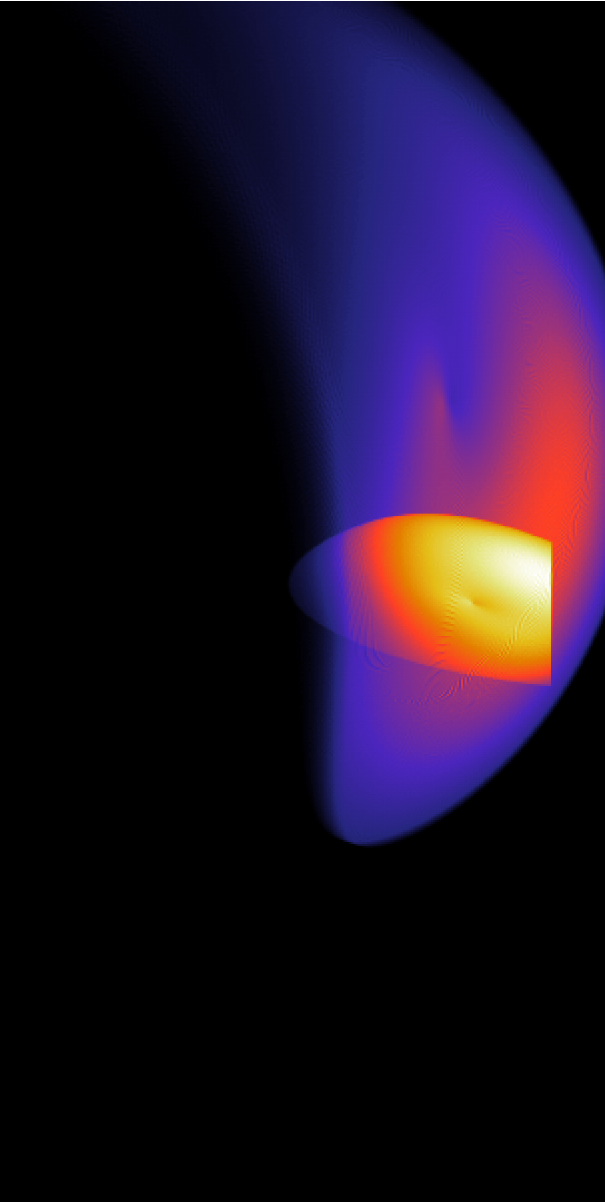}
        \put(4,91){\small\figlabelw{(d)}}
        \put(4,5){\small\figlabelw{$\tstar=\SI{0.7}{rad}$}}
    \end{overpic}
    \caption{
        Synchrotron radiation images generated with \SOFT\ for a few different
        values of the dominant particle pitch angle $\tstar$ in the magnetic
        equilibrium at $t=\SI{0.7344}{s}$. The dominant particle momentum is
        $\pstar = 50m_ec$.
    }
    \label{fig:paramscan}
\end{figure}

\SOFT\ simulations for the vertical scan in TCV discharge \#64614
are presented in figure~\ref{fig:softsynch}.  The synthetic images
are generated for approximately the same times as
figure~\ref{fig:expsynch}. \SOFT\ requires magnetic equilibria
at the desired times, and these were
obtained from experimental measurements using the Grad-Shafranov magnetic
reconstruction code \LIUQE~\cite{Moret2015}. From the
reconstructed equilibria, the relative distance between the plasma and
camera port were calculated as (a) $\Delta Z = \SI{9.3}{cm}$, (b)
$\Delta Z = \SI{4.3}{cm}$, (c) $\Delta Z = -\SI{1.7}{cm}$ and (d)
$\Delta Z = -\SI{3.4}{cm}$, respectively, in
figures~\ref{fig:expsynch} and~\ref{fig:softsynch}.

\newlength{\softimgwidth}
\setlength{\softimgwidth}{0.117\textwidth}
\begin{figure}
    \centering
    \begin{overpic}[width=\softimgwidth]{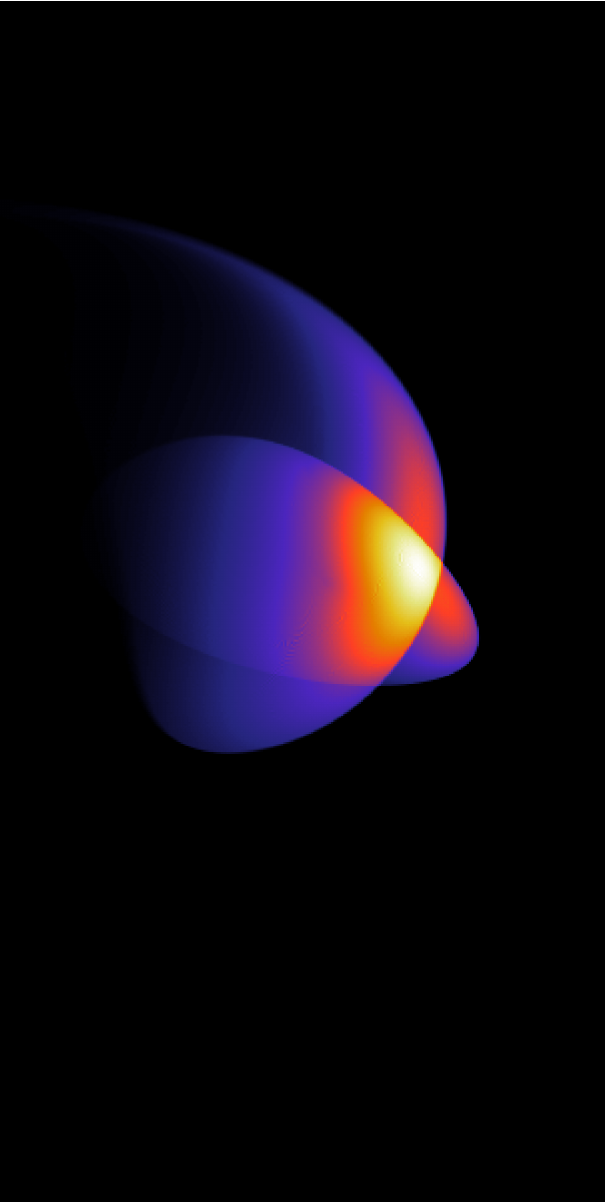}
        \put(4,91){\small\figlabelw{(a)}}
        \put(4,5){\scriptsize\figlabelw{$t=\SI{0.725}{s}$}}
    \end{overpic}
    \begin{overpic}[width=\softimgwidth]{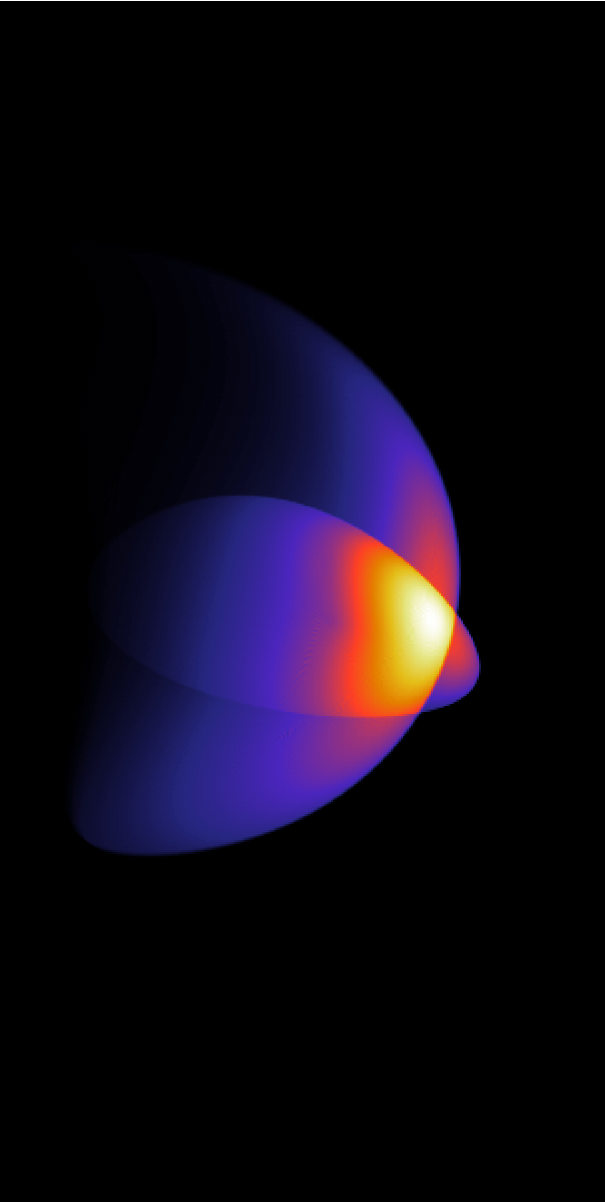}
        \put(4,91){\small\figlabelw{(b)}}
        \put(4,5){\scriptsize\figlabelw{$t=\SI{1.069}{s}$}}
    \end{overpic}
    \begin{overpic}[width=\softimgwidth]{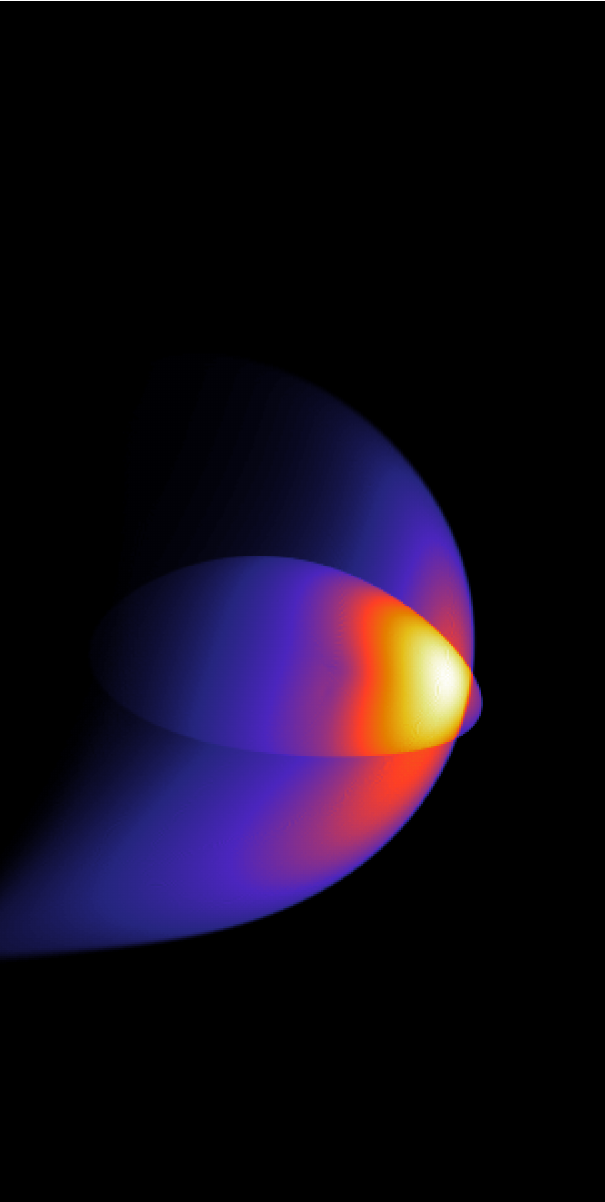}
        \put(4,91){\small\figlabelw{(c)}}
        \put(4,5){\scriptsize\figlabelw{$t=\SI{1.412}{s}$}}
    \end{overpic}
    \begin{overpic}[width=\softimgwidth]{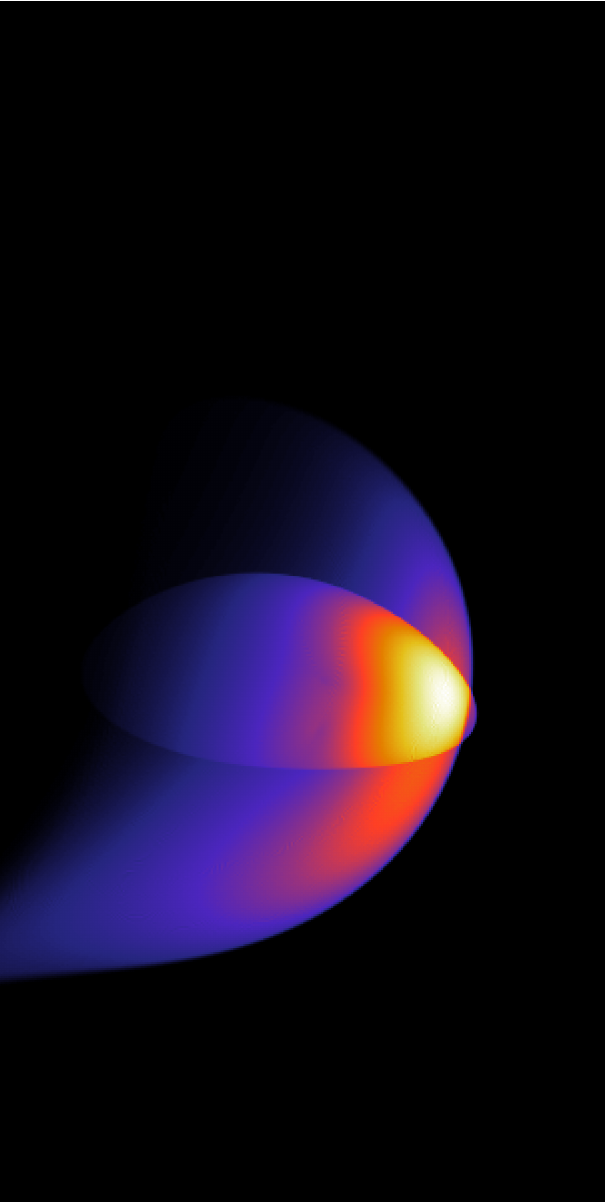}
        \put(4,91){\small\figlabelw{(d)}}
        \put(4,5){\scriptsize\figlabelw{$t=\SI{1.562}{s}$}}
    \end{overpic}
    \caption{
        Synthetic synchrotron images generated with \SOFT\ with reconstructed
        plasma equilibria from TCV discharge \#61614. In all simulations,
        $p^* = 50m_ec$ and $\tstar = \SI{0.4}{rad}$.
    }
    \label{fig:softsynch}
\end{figure}
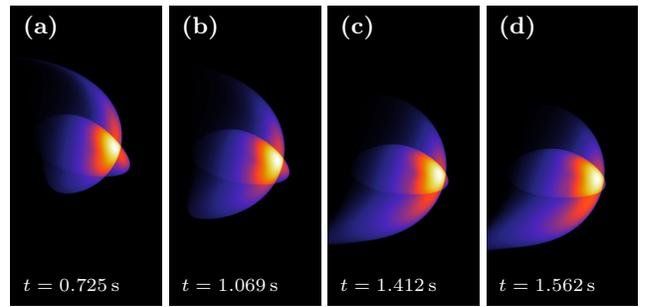

The simulations in figure~\ref{fig:softsynch} exhibit the same
dependence on $\Delta Z$ as the measurements in figure~\ref{fig:expsynch}.
As the plasma displaces down, the two parts of the synchrotron spot also
move down. The smaller spot also appears to move more slowly than the larger
spot, as described above. The underlying reason is the spatial origin of the
radiation: the smaller spot originates from a position far from the
detector, whereas the larger spot is located just in front of the detector,
i.e.\ the different rates are due to the observer's perspective.

The runaway parameters inferred from synchrotron imaging can also be compared
with predictions from conventional kinetic theory in axisymmetric magnetic
fields. The superthermal infinite-aspect ratio electron kinetic equation is
given by
\begin{equation*}
\!\frac{\partial f}{\partial \nu_c t}\! +\! \frac{E_\parallel}{E_c} \frac{\partial f}{\partial q_\parallel} = \frac{1}{q^2}\frac{\partial \gamma^2 f}{\partial q}\! +\! \frac{1\!+\!Z_\mathrm{eff}}{2}\frac{\gamma}{q^3}\frac{\partial}{\partial \xi}\!\left[(1-\xi^2)\frac{\partial f}{\partial \xi}\right] ,
\end{equation*}
where $\xi=\cos\theta$, $q= p/m_e c$, $\gamma = \sqrt{1+q^2}$, $Z_\mathrm{eff}$
the ion effective charge and $\nu_c = 4\pi \ln\Lambda n_e c r_0^2$ with $r_0$
the classical electron radius,  $\ln\Lambda$ the Coulomb logarithm. This kinetic
equation makes three predictions that are contradicted by the synchrotron
observations: (1) the primary (Dreicer) RE generation rate~\cite{Connor1975} at
$E/E_D = 6\%$ and $Z_\mathrm{eff}=1$ (assumed as no impurities were injected in
this flat-top runaway scenario) corresponds to a RE current-generation rate
$e c \partial n_\mathrm{RE}/\partial t \approx 1\,$GAm$^{-2}$s$^{-1}$; (2)
runaways with momentum $p \gg m_e c\sqrt{E_c/E} \approx 0.18m_e c$ will be
accelerated at a rate $\mathrm{d} p/\mathrm{d}t \approx eE_\parallel$
corresponding to a gain of $ 150\,m_e c$ per second---which due to significant
Dreicer generation would lead to a corresponding increase of $\pstar$; (3) at
relativistic speeds and small pitch angles, the runaway distribution takes the
form~\cite{Fulop2006} $f(t,\,p,\,\xi) = F(t,\,p)g(p,\,\xi)$ where the
pitch-angle distribution $g$ is independent of the evolving energy distribution
$F$, and is given by $g \approx \exp[-A(1-\xi)]$ where
$A = q(E/E_c)/(1+Z_\mathrm{eff})$. The dominant emitting pitch can be estimated
by weighing this pitch-angle distribution with the asymptotic
synchrotron-emission formula in the high-frequency limit~\cite{Bekefi}.

The result yields a maximum at
$\theta \approx  [(E/E_c)/(1+Z_\mathrm{eff})]^{1/3} m_e c/p \approx 8 m_e c/p$
which would correspond to $\theta_\star \approx 0.16$ at $p_\star=50\,m_e c$.
These considerations together indicate that the axisymmetric kinetic description
is inadequate for describing the RE dynamics in TCV discharge \#64614, a
conclusion that holds true also if measurement errors are taken into account.
A likely explanation is that magnetic perturbations play a significant role in the
dynamics. Low-frequency perturbations can transport REs out of the plasma, which
would suppress RE generation and maximum energies, whereas high-frequency
(kinetic) instabilities may cause pitch-angle
scattering~\cite{Aleynikov2015,Spong2018,Breizman2019review} which could explain
the anomalously high pitch angles. High-frequency kinetic instabilities have
previously been invoked to explain discrepancies between idealized kinetic
theory and experimental results under comparable plasma conditions in
DIII-D~\cite{Liu2018}.

Deviations in the appearance of the simulated synchrotron spots
compared to the experimentally measured spots should mainly be due to
(i) the particular choice~\eqref{eq:fREdelta} for the
distribution function, (ii) errors in the estimated dominant particle
$(\pstar, \tstar)$, (iii) errors in the calibrated detector and optical
distortion model, and (iv) errors in the magnetic measurements used for
reconstructing the plasma equilibrium. Error source (i) would primarily
affect the distribution of intensity across the synchrotron spot, and
if a full distribution function was used, it would give a more evenly
distributed and smooth radiation pattern. Errors in the dominant particle (ii)
primarily affect the size and overall (contour) shape of the spot,
while an error in detector position (iii) is expected to only have a slight
effect on the overall shape. Finally, errors in the plasma
equilibrium (iv) may affect both the observed synchrotron spot position,
its shape and the estimated dominant particle $(\pstar, \tstar)$.
For (i) and (ii), a sensitivity scan has been conducted, concluding that
the results presented here are robust, while for (iii) and (iv), estimates
suggests that errors are negligibly small.

    \section{Conclusions}\label{sec:conclusions}
    For the first time, the runaway electron synchrotron spot shape
    dependence on the vertical distance $\Delta Z$ between the runaway
    beam and the camera has been studied experimentally.
    It is found that the experimentally observed dependence qualitatively
    matches simulations conducted with the synthetic synchrotron
    diagnostic \SOFT\ well. These experiments therefore validate an important
    geometrical aspect of the theory underlying the synthetic
    diagnostic, and lend confidence to its capability of describing
    RE radiation in tokamaks. The
    multiple vertical views did however not provide sufficient
    additional information to more accurately constrain the runaway electron
    distribution function. This is primarily due to the large dominant
    pitch angle of the observed synchrotron spot, which allows the
    full vertical extent of the runaway beam to be observed regardless
    of the vertical position of the detector.

    When comparing observations to predictions of conventional kinetic
    runaway theory, we find that kinetic theory significantly
    underestimate the observed dominant pitch angle $\tstar$ of the
    particles, while simultaneously greatly overestimating their energies.
    We hypothesise that the anomalously large dominant pitch angle observed
    could be due to the presence of kinetic instabilities, although no diagnostic
    was installed during the experiment to confirm this hypothesis. The runaway
    energies are most likely overestimated due to significant radial transport.

    The ability of the \MultiCam\ system to simultaneously obtain several images
    at different wavelengths was not utilised in this work, as it was not needed
    for the synthetic diagnostic validation. However, synchrotron radiation
    images at multiple wavelengths are expected to provide complementary information
    that can be used to further constrain the distribution function, and provide
    additional data points for validation of kinetic theory. This possibility will
    be explored in a future publication.

    \ack
    The authors gratefully acknowledge the help of A. Tema Biwole with retrieving
    plasma parameter signals.
    This work has been carried out within the framework of the EUROfusion Consortium and has received funding from the Euratom research and training programme 2014 - 2018 and 2019 - 2020 under grant agreement No 633053. This project has received funding from the European Research Council (ERC) under the European Union's Horizon 2020 research and innovation programme under grant agreement No 647121. The views and opinions expressed herein do not necessarily reflect those of the European Commission.  The work was also supported by the Swedish Research Council (Dnr.~2018-03911) and the EUROfusion - Theory and Advanced Simulation Coordination (E-TASC). 
    \section*{References}
    \bibliography{ref}
    
\end{document}